\pdfoutput=1
\RequirePackage{ifpdf}
\ifpdf 
\documentclass[pdftex]{sigma}
\else
\documentclass{sigma}
\fi

\begin{document}

\allowdisplaybreaks

\renewcommand{\PaperNumber}{045}

\FirstPageHeading

\numberwithin{equation}{section}

\newtheorem{thm}{Theorem}[section]
\newtheorem{Mthm}{Main Theorem}
\newtheorem{prop}[thm]{Proposition}
\newtheorem{cor}[thm]{Corollary}
\newtheorem{cjt}[thm]{Conjecture} \newtheorem{lem}[thm]{Lemma}

{ \theoremstyle{definition}
\newtheorem{de}[thm]{Definition}
\newtheorem{rem}[thm]{Remark}
\newtheorem{ex}[thm]{Example}
}

\newcommand{\p}{\partial}
\def\la{\langle}
\def\ra{\rangle}
\def\dsum{\displaystyle\sum}

\ShortArticleName{Euler Equations Related to the Generalized Neveu--Schwarz Algebra}

\ArticleName{Euler Equations Related
\\
to the Generalized Neveu--Schwarz Algebra}

\Author{Dafeng ZUO~$^{\dag\ddag}$}

\AuthorNameForHeading{D.~Zuo}

\Address{$^\dag$~School of Mathematical Science, University of Science and Technology of China,
\\
\hphantom{$^\dag$}~Hefei 230026, P.R.~China} \EmailD{\href{dfzuo@ustc.edu.cn}{dfzuo@ustc.edu.cn}}

\Address{$^\ddag$~Wu Wen-Tsun Key Laboratory of Mathematics, USTC,
\\
\hphantom{$^\ddag$}~Chinese Academy of Sciences, P.R.~China}

\ArticleDates{Received March 11, 2013, in f\/inal form June 12, 2013; Published online June 16, 2013}

\Abstract{In this paper, we study supersymmetric or bi-superhamiltonian Euler equations related to the
generalized Neveu--Schwarz algebra.
As an application, we obtain several supersymmetric or bi-superhamiltonian generalizations of some
well-known integrable systems including the coupled KdV equation, the 2-component Camassa--Holm equation
and the 2-component Hunter--Saxton equation.
To our knowledge, most of them are new.}

\Keywords{supersymmetric; bi-superhamiltonian; Euler equations; generalized Neveu--Schwarz algebra}

\Classification{37K10; 35Q51}

\section{Introduction}

For a~classical rigid body with a~f\/ixed point, the conf\/iguration space is the group ${\rm SO}(3)$ of
rotations of three-dimensional Euclidean space.
In 1765, L.~Euler proposed the equations of motion of the rigid body describing as geodesics in ${\rm
SO}(3)$, where ${\rm SO}(3)$ is provided with a~left-invariant metric.
In essence, the Euler theory of a~rigid theory is fully described by this invariance.

Let $G$ be an arbitrary (possibly inf\/inite-dimensional) Lie group and $\mathcal{G}$ the corresponding Lie
algebra and $\mathcal{G}^{*}$ the dual of $\mathcal{G}$.
V.I.~Arnold in~\cite{Arn1966} suggested a~general framework for Euler equations on~$G$, which can be
regarded as a~conf\/iguration space of some physical systems.
In this framework Euler equations describe geodesic f\/lows w.r.t.\ suitable one-side invariant Riemannian
metrics on~$G$ and can be given to a~variety of conservative dynamical systems in mathematical physics, for
instance, see~\cite{Ara2007,AK1998,
Con2007,Con2003,Con2006,DS2001,EM1970,Guha2008,Guha2000,Guha2006,K2005,BLG2008,KM2003,KW2009,Kolev2007,JO2008,JO2009,Man1976,Marsden,Mish1970,M1998,OK1987,Ra1981,Zuo-2010-1}
and references therein.

Since V.~Ovsienko and B.~Khesin in~\cite{OK1987} interpreted the Kuper--KdV equation~\cite{Kuper1984} as
a~geodesic f\/low equation on the superconformal group w.r.t.\ an $L^2$-type metric, it has been attracted
a~lot of interest in studying super (fermionic or supersymmetric) anologue of Arnold's approach, which has
some dif\/ferent characteristic f\/lavors, for instance~\cite{Ara2007,DS2001,Guha2006,
Kuper1984,JO2008, JO2009,PO2008, ZZ2011}.

In this paper, we are interested in Euler equations related to the $N=1$ generalized Neveu--Schwarz (GNS in
brief) algebra {\it $\mathcal{G}$}, which was introduced by P.~Marcel, V.~Ovsienko and C.~Roger
in~\cite{MOR1997} as a~generalization of the $N=1$ Neveu--Schwarz algebra and the extended Virasoro algebra.
In~\cite{Guha2006}, P.~Guha and P.J.~Ovler have studied the Euler equations related to the GNS algebra
$\mathcal{G}$ and obtained fermionic versions of the 2-component Camassa--Holm equation and the Ito
equation in some special metrics.
Our motivations are twofold.
One is to study the Euler equation related to~$\mathcal{G}$ for a~more general metric
$M_{c_1,c_2,c_3,c_4,c_5,c_6}$ in~\eqref{eq2.2} with six-parameters given by
\begin{gather*}
\la\hat{F},\hat{G}\ra=\int_{S^1}\big(c_1fg+c_2f_xg_x+c_3\phi\p^{-1}\chi+c_4\phi_x\chi+c_5ab+c_6\alpha\p^{-1}
\beta\big)dx+\vec{\sigma}\cdot\vec{\tau},
\end{gather*}
which can be regarded as a~super-version of Sobolev-metrics in the super space.
The other is to study the condition under which Euler equations are supersymmetric or bi-superhamiltonian.
Our main results is to show that
\begin{gather*}
\begin{tabular}{|c|c|c|}
 \hline
$\diagdown$ the Euler equation is & bi-superhamiltonian & supersymmetric \\
when the metric is $\diagdown$&~&\\
\hline
 $M_{c_1,c_2,\frac{1}{4}c_1,c_2,c_5,-c_5}$ & Yes & No (if $c_1\ne 0$)\bsep{4pt} \\
 \hline
 $M_{c_1,c_2,c_1,c_2,c_5,-c_5}$ & only f\/ind a superhamiltonian
 & Yes \\
 ~&structure (if $c_1\ne 0$) &~\\
 \hline
 $M_{0,c_2,0,c_2,c_5,-c_5}$ & Yes & Yes \\
 \hline
\end{tabular}
\end{gather*}
As a~byproduct, we obtain some supersymmetric or bi-superhamiltonian generalizations of some well-known
integrable systems including the coupled KdV equation, the 2-component Camassa--Holm equation and the
2-component Hunter--Saxton equation.

This paper is organized as follows.
In Section~\ref{section2}, we calculate the Euler equation on $\mathcal{G}_{\rm reg}^{*}$ and discuss their
Hamiltonian properties.
In Section~\ref{section3}, we study bi-superhamiltonian Euler equations.
Section~\ref{section4} is devoted to describe supersymmetric Euler equations, also including a~class of
both supersymmetric and bi-superhamiltonian Euler equations.
A few concluding remarks are given in the last section.

\section{Euler equations related to the GNS algebra}
\label{section2}

To be self-contained, let us recall the Anorld's approach~\cite{AK1998,KM2003,KW2009}.
Let $G$ be an arbitrary Lie group and $\mathcal{G}$ the corresponding Lie algebra and~$\mathcal{G}^{*}$ the
dual of $\mathcal{G}$.
Firstly let us f\/ix a~energy quadratic form $E(v)=\frac{1}{2}\la v,\mathcal{A}v\ra^{*}$ on~$\mathcal{G}$ and
consider right translations of this quadratic form on~$\mathcal{G}$.
Then the energy quadratic form def\/ines a~right-invariant Riemannian metric on~$G$.
The geodesic f\/low on~$G$ w.r.t.\ this energy metric represents the extremals of the least action
principle, i.e., the actual motions of our physical system.
For a~rigid body, one has to consider left translations.
We next identify $\mathcal{G}$ and its dual $\mathcal{G}^{*}$ with the help of $E(\cdot)$.
This identif\/ication $\mathcal{A}:\mathcal{G}\to\mathcal{G}^{*}$, called an inertia operator, allows us to
rewrite the Euler equation on~$\mathcal{G}^{*}$.
It turns out that the Euler equation on~$\mathcal{G}^{*}$ is Hamiltonian w.r.t.\ a canonical Lie--Poisson
structure on~$\mathcal{G}^{*}$.
Notice that in some cases it turns out to be not only Hamiltonian, but also bihamiltonian.
Moreover, the corresponding Hamiltonian function is $-E(m)=-\frac{1}{2}\la\mathcal{A}^{-1}m,m\ra^{*}$ lifted
from the Lie algebra $\mathcal{G}$ to its dual space~$\mathcal{G}^{*}$, where
$m=\mathcal{A}v\in\mathcal{G}^{*}$.
\begin{de}[\cite{AK1998,KM2003}]
\label{de1.1}
The Euler equation on $\mathcal{G}^{*}$, corresponding to the right-invariant metric
$-E(m)=-\frac{1}{2}\la\mathcal{A}^{-1}m,m\ra^{*}$ on $G$, is given by the following explicit formula
\begin{gather*}
\frac{dm}{dt}=-{\rm ad}^{*}_{\mathcal{A}^{-1}m}m,
\end{gather*}
as an evolution of a~point $m\in\mathcal{G}^{*}$.
\end{de}

In the following, we take $\mathcal{G}$ to be the $N=1$ generalized Neveu--Schwarz algebra~\cite{OV1995}.
Let $V$ be a~$\mathbb{Z}_2$ graded vector space, i.e., $V=V_B\oplus V_F$.
An element $v$ of $V_B$ (resp., $V_F$) is said to be even (resp., odd).
The super commutator of a~pair of elements $v,w\in V$ is def\/ined by
\begin{gather*}
[v,w]=vw-(-1)^{|v||w|}wv.
\end{gather*}

Let $\mathcal{D}^s\big(S^1\big)$ be the group of orientation preserving Sobolev $H^s$ dif\/feomorphisms of the
circle and $T_{id}\mathcal{D}^s\big(S^1\big)$ the corresponding Lie algebra of vector f\/ields, denoted by
$\hbox{Vect}^s\big(S^1\big)=\big\{f(x)\frac{d}{dx}|f(x)\in H^s\big(S^1\big)\big\}$.
We denote
\begin{gather*}
V_B=\hbox{Vect}^s\big(S^1\big)\oplus\hbox{C}^\infty\big(S^1\big)\oplus\mathbb{R}^3,
\qquad
V_F=\hbox{C}^\infty\big(S^1\big)\oplus\hbox{C}^\infty\big(S^1\big).
\end{gather*}
\begin{de}[\cite{OV1995}] The GNS algebra $\mathcal{G}$ is an algebra $V_B\oplus V_F$ with the commutation relation
given by
\begin{gather*}
[\hat{F},\hat{G}]=\left(\left(fg_x-f_xg+\frac{1}{2}\phi\chi\right)\frac{d}{dx},
 \left(f\chi_x-\frac{1}{2}f_x\chi-g\phi_x+\frac{1}{2}g_x\phi\right)dx^{-\frac{1}{2}},\right.
\\
\left. \hphantom{[\hat{F},\hat{G}]=}{}
fb_x-a_xg+\frac{1}{2}\phi\beta+\frac{1}{2}\alpha\chi,
\left(f\beta_x+\frac{1}{2}f_x\beta-\frac{1}{2}a_x\chi-g\alpha_x-\frac{1}{2}g_x\alpha+\frac{1}{2}
b_x\phi\right)dx^\frac{1}{2},
\vec{\omega}\right),
\end{gather*}
where $\phi$, $\chi$, $\alpha$ and $\beta$ are fermionic functions, and $f$, $g$, $a$ and $b$ are bosonic
functions, and
$\hat{F}=\big(f(x,t)\frac{d}{dx},\phi(x,t)dx^{-\frac{1}{2}},a(x,t),\alpha(x,t)dx^{\frac{1}{2}},\vec{\sigma}\big)\in\mathcal{G}$ and $\hat{G}=\big(g(x,t)\frac{d}{dx},\chi(x,t)dx^{-\frac{1}{2}},b(x,t)$, $\beta(x,t)dx^{\frac{1}{2}}, \vec{\tau}\big)\in\mathcal{G}$ and $\vec{\sigma},\vec{\tau}\in\mathbb{R}^3$ and $\vec{\omega}=(\omega_1,\omega_2,\omega_3)\in\mathbb{R}^3$. Here
\begin{gather*}
\omega_1(\hat{F},\hat{G})=\int_{S^1}(f_xg_{xx}+\phi_x\chi_x)dx,
\\
\omega_2(\hat{F},\hat{G})=\int_{S^1}(f_{xx}b-g_{xx}a-\phi_x\beta+\chi_x\alpha)dx,
\\
\omega_3(\hat{F},\hat{G})=\int_{S^1}(2ab_{x}+2\alpha\beta)dx.
\end{gather*}
\end{de}

Let us denote
\begin{gather*}
{\mathcal{G}}_{\rm reg}^{*}=\hbox{C}^\infty\big(S^1\big)\oplus\hbox{C}^\infty\big(S^1\big)\oplus\hbox{C}
^\infty\big(S^1\big)\oplus\hbox{C}^\infty\big(S^1\big)\oplus\mathbb{R}^3
\end{gather*}
to be the regular part of the dual space ${\mathcal{G}}^{*}$ to ${\mathcal{G}}$, under the following pair
\begin{gather*}
\la\hat{U},\hat{F}\ra^{*}=\int_{S^1}(u f+\psi\phi+va+\gamma\alpha)dx+\vec{\varsigma}\cdot\vec{\sigma},
\end{gather*}
where
$\hat{U}=\big(u(x,t)dx^2,\psi(x,t)dx^{\frac{3}{2}},v(x,t)dx,\gamma(x,t)dx^\frac{1}{2},\vec{\varsigma}\big)\in{\mathcal{G}}^{*}$ and $\vec{\varsigma}=(\varsigma_1,\varsigma_2,\varsigma_3)\in\mathbb{R}^3$.
By the def\/inition, using integration by parts we have
\begin{gather*}
\la {\rm ad}^{*}_{\hat{F}}(\hat{U}),\hat{G}\ra^{*}=-\la\hat{U},[\hat{F},\hat{G}]\ra^{*} =-\int_{S^1}\left(u\left(fg_x-f_xg+\frac{1}{2}\phi\chi\right)\right.
\\
\qquad\quad{}
+\psi\left(f\chi_x-\frac{1}{2}f_x\chi-g\phi_x+\frac{1}{2}
g_x\phi\right)+v\left(fb_x-a_xg+\frac{1}{2}\phi\beta+\frac{1}{2}\alpha\chi\right)
\\
\left.
\qquad\quad{}
+\gamma\left(f\beta_x+\frac{1}{2}f_x\beta-\frac{1}{2}
a_x\chi-g\alpha_x-\frac{1}{2}g_x\alpha+\frac{1}{2}b_x\phi\right)\right)dx-\vec{\varsigma}\cdot\vec{\omega}
\\
\qquad{}
=\int_{S^1}\left(2uf_x+u_xf-\varsigma_1f_{xxx}+\varsigma_2a_{xx}+\frac{3}{2}\psi\phi_x+\frac{1}{2}
\psi_x\phi+\frac{1}{2}\gamma\alpha_x-\frac{1}{2}\gamma_x\alpha+va_x\right)gdx
\\
\qquad\quad{}
+\int_{S^1}\left({\varsigma}_1\phi_{xx}-{\varsigma}_2\alpha_x-\frac{1}{2}u\phi-\frac{1}{2}v\alpha+\frac{3}
{2}f_x\psi+f\psi_x+\frac{1}{2}\gamma a_x\right)\chi dx
\\
\qquad\quad{}
+\int_{S^1}\left((vf)_x+\frac{1}{2}(\gamma\phi)_x-\varsigma_2f_{xx}+2\varsigma_3a_x\right)bdx
\\
\qquad\quad{}
+\int_{S^1}\left(\gamma_xf+\frac{1}{2}\gamma f_x-\frac{1}{2}
v\phi+\varsigma_2\phi_x-2\varsigma_3\alpha\right)\beta dx.
\end{gather*}
So the coadjoint action on $\mathcal{G}_{\rm reg}^{*}$ is given by
\begin{gather*}
{\rm ad}^{*}_{\hat{F}}(\hat{U})=\left(\left(2uf_x+u_xf-\varsigma_1f_{xxx}+\varsigma_2a_{xx}+\frac{3}{2}
\psi\phi_x+\frac{1}{2}\psi_x\phi+\frac{1}{2}\gamma\alpha_x-\frac{1}{2}\gamma_x\alpha+va_x\right){dx^2},\right.
\\
\hphantom{{\rm ad}^{*}_{\hat{F}}(\hat{U})=}{}
\left({\varsigma}_1\phi_{xx}-{\varsigma}_2\alpha_x-\frac{1}{2}u\phi-\frac{1}{2}v\alpha+\frac{3}{2}
f_x\psi+f\psi_x+\frac{1}{2}\gamma a_x\right)dx^{\frac{3}{2}},
\\
\left.
\hphantom{{\rm ad}^{*}_{\hat{F}}(\hat{U})=}{}
\left(\! (vf)_x\!+\frac{1}{2}(\gamma\phi)_x\!-\varsigma_2f_{xx}\!+2\varsigma_3a_x\right)\!dx,\!
\left(\!\gamma_xf\!+\frac{1}{2}\gamma f_x\!-\frac{1}{2}
v\phi\!+\varsigma_2\phi_x\!-2\varsigma_3\alpha\right)\!dx^{\frac{1}{2}},
0\!\right)\!.
\end{gather*}

On $\mathcal{G}$, let us introduce an inner product $M_{c_1,c_2,c_3,c_4,c_5,c_6}$ given by
\begin{gather}
\la\hat{F},\hat{G}\ra=\int_{S^1}\big(c_1fg+c_2f_xg_x+c_3\phi\p^{-1}\chi+c_4\phi_x\chi+c_5ab+c_6\alpha\p^{-1}
\beta\big)dx+\vec{\sigma}\cdot\vec{\tau},
\label{eq2.2}
\end{gather}
which is a~generalization of that in~\cite{DS2001,Guha2006}.
By the Def\/inition~\ref{de1.1}, the Euler equation on $\mathcal{G}_{\rm reg}^{*}$ for
$M_{c_1,c_2,c_3,c_4,c_5,c_6}$ is
\begin{gather}
\frac{d\hat{U}}{dt}=-{\rm ad}^{*}_{\mathcal{A}^{-1}\hat{U}}\hat{U}
\label{eq2.4}
\end{gather}
as an evolution of a~point
$\hat{U}\!=\!\big(u(x,t)dx^2,\psi(x,t)dx^{\frac{3}{2}},v(x,t),\gamma(x,t)dx^\frac{1}{2},\vec{\varsigma}\big)\!\in\!{\mathcal{G}}^{*}$, where $\mathcal{A}:\mathcal{G}\!\to\! \mathcal{G}^{*}$ is an inertia operator def\/ined by
\begin{gather*}
\la\hat{F},\hat{G}\ra=\la\mathcal{A}(\hat{F}),\hat{G}\ra^{*}.
\end{gather*}
A direct computation shows that the inertia operator $\mathcal{A}:\mathcal{G}\to \mathcal{G}^{*}$
has the form
\begin{gather*}
\mathcal{A}(\hat{F})=\big(\Lambda(f)dx^2,\Theta(\phi)dx^{\frac{3}{2}},c_5adx,c_6\p^{-1}\alpha dx^{\frac{1}
{2}},\vec{\sigma}\big),
\end{gather*}
where $\Lambda(f)=c_1f-c_2f_{xx}$ and $\Theta(\phi)=c_4\phi_x-c_3\p^{-1}\phi$.
Thus we have
\begin{prop}
The Euler equation~\eqref{eq2.4} on $\mathcal{G}_{\rm reg}^{*}$ for $M_{c_1,c_2,c_3,c_4,c_5,c_6}$ reads
\begin{gather}
u_t=\varsigma_1f_{xxx}-\varsigma_2a_{xx}-2uf_x-u_xf-va_x-\frac{3}{2}\psi\phi_x-\frac{1}{2}
\psi_x\phi-\frac{1}{2}\gamma\alpha_x,\nonumber
\\
\psi_t=\frac{1}{2}u\phi+\frac{1}{2}v\alpha-{\varsigma}_1\phi_{xx}+{\varsigma}_2\alpha_x-\frac{3}{2}
f_x\psi-f\psi_x-\frac{1}{2}\gamma a_x,\nonumber
\\
v_t=\varsigma_2f_{xx}-2\varsigma_3a_x-(vf)_x-\frac{1}{2}(\gamma\phi)_x,
\label{eq2.6}
\\
\gamma_t=\frac{1}{2}v\phi-\gamma_xf-\frac{1}{2}\gamma f_x-\varsigma_2\phi_x+2\varsigma_3\alpha,\nonumber
\end{gather}
where $u=\Lambda(f)=c_1f-c_2f_{xx}$, $\psi=\Theta(\phi)=c_4\phi_x-c_3\p^{-1}\phi$, $v=c_5a$ and
$\gamma=c_6\p^{-1}\alpha$.
\end{prop}

Let us remark that the system~\eqref{eq2.6} has been obtained in~\cite{Guha2006} with minor typos.
But they didn't discuss the condition under which the Euler equation~\eqref{eq2.6} is supersymmetric or
bi-superhamiltonian.

According to Def\/inition~\ref{de1.1}, the Euler equation~\eqref{eq2.6} has a~natural Hamiltonian
description~\cite{AK1998,KM2003,KW2009}.
Let $F_i:\mathcal{G}^{*}\to\mathbb{R}$, $i=1,2$, be two arbitrary smooth functionals.
The dual space $\mathcal{G}^{*}$ carries a~canonical Lie--Poisson bracket
\begin{gather*}
\{F_1,F_2\}_2(\hat{U})=\left\la\hat{U},\left[\frac{\delta F_1}{\delta\hat{U}},\frac{\delta F_2}{\delta\hat{U}}\right]\right\ra^{*},
\end{gather*}
where $\hat{U}\in\mathcal{G}^{*}$ and $\frac{\delta F_i}{\delta\hat{U}}=\big(\frac{\delta F_i}{\delta
u},\frac{\delta F_i}{\delta\psi},\frac{\delta F_i}{\delta v},\frac{\delta F_i}{\delta\gamma},\frac{\delta
F_i}{\delta\vec{\varsigma}}\big)\in\mathcal{G},i=1,2$.
The induced superhamiltonian operator is given by
\begin{gather}
\mathcal{J}_2=\left(
\begin{matrix}
\varsigma_1\p^3-u\p-\p u&-\psi\p-\frac{1}{2}\p\psi,&-\varsigma_2\p^2-v\p&-\gamma\p+\frac{1}{2}\p\gamma
\\
-\p\psi-\frac{1}{2}\psi\p&\frac{1}{2}u-\varsigma_1\p^2&-\frac{1}{2}\gamma\p&\frac{1}{2}v+\varsigma_2\p
\\
\varsigma_2\p^2-\p v&-\frac{1}{2}\p\gamma&-2\varsigma_3\p&0
\\
-\p\gamma+\frac{1}{2}\gamma\p&\frac{1}{2}v-\varsigma_2\p&0&2\varsigma_3
\end{matrix}
\right).
\label{eq2.8}
\end{gather}
\begin{prop}
The Euler equation~\eqref{eq2.6} could be written as
\begin{gather}
\frac{d}{dt}(u,\psi,v,\gamma)^{\rm T}=\mathcal{J}_2\left(\frac{\delta H_1}{\delta u},\frac{\delta H_1}
{\delta\psi},\frac{\delta H_1}{\delta v},\frac{\delta H_1}{\delta\gamma}\right)^{\rm T}
\label{eq2.9}
\end{gather}
with the Hamiltonian $H_1=\frac{1}{2}\int_{S^1}(uf+\psi\phi+va+\gamma\alpha)dx$, where $(\cdot )^{\rm T}$ means the
transpose of vectors.
\end{prop}
\begin{proof}
Indeed, for a~functional $F[u,\psi,v,\gamma]$, the variational derivatives $\frac{\delta F}{\delta u}$,
$\frac{\delta F}{\delta\psi}$, $\frac{\delta F}{\delta v}$ and $\frac{\delta F}{\delta\gamma}$ are
def\/ined by
\begin{gather}
\frac{d}{d\epsilon}\mid_{\epsilon=0}
F[u+\epsilon\delta u,\psi+\epsilon\delta\psi,v+\epsilon\delta v,\gamma+\epsilon\delta\gamma]\nonumber
\\
\qquad{} =\int\left(\delta u\frac{\delta F}{\delta u}+\delta\psi\frac{\delta F}{\delta\psi}+\delta v\frac{\delta F}
{\delta v}+\delta\gamma\frac{\delta F}{\delta\gamma}\right)dx.
\label{eq2.13}
\end{gather}
By using~\eqref{eq2.13}, we have
\begin{gather*}
\frac{\delta H_1}{\delta f}=\Lambda(f),
\qquad
\frac{\delta H_1}{\delta\phi}=-\Theta(\phi),
\qquad
\frac{\delta H_1}{\delta a}=v,
\qquad
\frac{\delta H_1}{\delta\alpha}=-\gamma.
\end{gather*}
It follows from the def\/inition of $u$, $\psi$ and $\gamma$ that
\begin{gather}
\frac{\delta H_1}{\delta u}=\Lambda^{-1}\frac{\delta H_1}{\delta u}=f,
\qquad
\frac{\delta H_1}{\delta\psi}=-\Theta^{-1}\frac{\delta H_1}{\delta\phi}=\phi,\nonumber
\\
\frac{\delta H_1}{\delta v}=a,
\qquad
\frac{\delta H_1}{\delta\gamma}=-\p\frac{\delta H_1}{\delta\alpha}=\alpha.\label{eq2.14}
\end{gather}
Hence,~\eqref{eq2.9} could be easily verif\/ied by using~\eqref{eq2.8} and~\eqref{eq2.14}.
\end{proof}

\section[Bihamiltonian Euler equations on $\mathcal{G}_{\rm reg}^{*}$]{Bihamiltonian Euler equations on
$\boldsymbol{\mathcal{G}_{\rm reg}^{*}}$}
\label{section3}

{\sloppy Unless otherwise stated, in the following we use ``(bi)hamiltonian'' to denote ``(bi)-super\-hamil\-tonian".
In this section we want to study bihamiltonian Euler equations on $\mathcal{G}_{\rm reg}^{*}$ w.r.t.\ the
metric $M_{c_1,c_2,\frac{1}{4}c_1,c_2,c_5,-c_5}$ and propose some new bihamiltonian and fermionic
extensions of well-known integrable systems including coupled the KdV equation, the 2-CH equation and the 2-HS
equation.

}

\subsection[The frozen Lie--Poisson bracket on $\mathcal{G}_{\rm reg}^{*}$]{The frozen Lie--Poisson bracket
on $\boldsymbol{\mathcal{G}_{\rm reg}^{*}}$}

For the purpose of discussing possible bihamiltonian Euler equations, we introduce a~frozen Lie--Poisson
bracket on $\mathcal{G}_{\rm reg}^{*}$ def\/ined by
\begin{gather*}
\{F_1,F_2\}_1(\hat{U})=\left\la\hat{U}_0,\left[\frac{\delta F_1}{\delta\hat{U}},\frac{\delta F_2}{\delta\hat{U}}
\right]\right\ra^{*},
\end{gather*}
for a~f\/ixed point $\hat{U}_0\in{\mathcal{G}}^{*}$.
The corresponding Hamiltonian equation is given by
\begin{gather}
\frac{d\hat{U}}{dt}=-{\rm ad}^{*}_{\frac{\delta H_2}{\delta\hat{U}}}\hat{U}_0
\label{eq2.10}
\end{gather}
for a~functional $H_2:{\mathcal{G}^{*}_{\rm reg}}\to\mathbb{R}$.
If we could f\/ind a~functional $H_2$ and a~suitable point $\hat{U}_0\in\mathcal{G}_{\rm reg}^{*}$ such that
the system~\eqref{eq2.10} coincides with~\eqref{eq2.6}.
This means that the Euler equation~\eqref{eq2.6} is bihamiltonian and could be written as
\begin{gather*}
\frac{d}{dt}(u,\psi,v,\gamma)^{\rm T}=\mathcal{J}_1\left(\frac{\delta H_2}{\delta u},\frac{\delta H_2}
{\delta\psi},\frac{\delta H_2}{\delta v},\frac{\delta H_2}{\delta\gamma}\right)^{\rm T}=\mathcal{J}
_2\left(\frac{\delta H_1}{\delta u},\frac{\delta H_1}{\delta\psi},\frac{\delta H_1}{\delta v},\frac{\delta H_1}
{\delta\gamma}\right)^{\rm T}
\end{gather*}
with Hamiltonian operators $\mathcal{J}_2$ in~\eqref{eq2.8} and
$\mathcal{J}_1=\mathcal{J}_2|_{\hat{U}=\hat{U}_0}$.
Moreover, according to Proposition~5.3 in~\cite{KM2003}, $\{~,~\}_1$ and $\{~,~\}_2$ are compatible for
every freezing point $\hat{U}_0$.

\subsection[Bihamiltonian Euler equations on $\mathcal{G}_{\rm reg}^{*}$ w.r.t.\
$M_{c_1,c_2,\frac{1}{4}c_1,c_2,c_5,-c_5}$]{Bihamiltonian Euler equations on $\boldsymbol{\mathcal{G}_{\rm reg}^{*}}$ w.r.t.\
$\boldsymbol{M_{c_1,c_2,\frac{1}{4}c_1,c_2,c_5,-c_5}}$}

In this case, we have
\begin{gather*}
c_1=4\,c_3,
\qquad
c_2=c_4,
\qquad
c_6=-c_5.
\end{gather*}
By setting $\phi=\eta_x$ and $\alpha=\mu_x$, then
\begin{gather}
u=\Lambda(f)=c_1f-c_2f_{xx},
\qquad
\psi=\Pi(\eta)=c_2\eta_{xx}-\frac{1}{4}c_1\eta,
\qquad
v=c_5a,
\qquad
\gamma=-c_5\mu
\label{eq4.0}
\end{gather}
and the Euler equation becomes
\begin{gather}
u_t=\varsigma_1f_{xxx}-\varsigma_2a_{xx}-2uf_x-u_xf-\frac{3}{2}\psi\eta_{xx}-\frac{1}{2}
\psi_x\eta_x-va_x-\frac{1}{2}\gamma\mu_{xx},\nonumber
\\
\psi_t=\frac{1}{2}u\eta_x-{\varsigma}_1\eta_{xxx}+{\varsigma}_2\mu_{xx}-\frac{3}{2}
f_x\psi-f\psi_x+\frac{1}{2}(v\mu)_x,\nonumber
\\
v_t=\varsigma_2f_{xx}-2\varsigma_3a_x-(vf)_x-\frac{1}{2}(\gamma\eta_x)_x,
\label{eq4.1}
\\
\gamma_t=\frac{1}{2}v\eta_x-\gamma_xf-\frac{1}{2}\gamma f_x-\varsigma_2\eta_{xx}+2\varsigma_3\mu_x.\nonumber
\end{gather}
We are now in a~position to state our main theorem.
\begin{thm}
\label{thm3.1}
The system~\eqref{eq4.1} is bihamiltonian on $\mathcal{G}_{\rm reg}^{*}$ with a~freezing point
$\hat{U}_0=\big(\frac{c_1}{2}dx^2,0$, $0,0,\big(c_2,0,\frac{c_5}{2}\big)\big)\in\mathcal{G}_{\rm reg}^{*}$ and
a~Hamiltionian functional
\begin{gather*}
H_2=\int_{S^1}\left({-}\frac{\varsigma_1}{2}ff_{xx}+\frac{c_1}{2}f^3
-\frac{c_2}{4}f^2f_{xx}-\frac{c_2}{2}f\eta_x\eta_{xx}-\frac{3c_1}{8}f\eta\eta_x+\frac{\varsigma_1}{2}
\eta\eta_{xx}\right.\nonumber
\\
\left.\hphantom{H_2=}{}
-\varsigma_2af_x+\frac{1}{2}avf+\varsigma_3a^2+\frac{1}{2}a\gamma\eta_x+\varsigma_3\mu\mu_{x}
+\varsigma_2\mu\eta_{xx}+\frac{1}{2}\gamma\mu_xf
\right)dx.\nonumber
\end{gather*}
\end{thm}
\begin{proof}
Direct computation gives
\begin{gather}
\frac{\delta H_2}{\delta f}=\varsigma_2a_x-\varsigma_1f_{xx}+\frac{3c_1}{2}f^2-c_2ff_{xx}-\frac{c_2}{2}
f_x^2
-\frac{c_2}{2}\eta_x\eta_{xx}-\frac{3c_1}{8}\eta\eta_{xx}+\frac{1}{2}av+\frac{1}{2}\gamma\mu_x,\nonumber
\\
\frac{\delta H_2}{\delta\eta}=\frac{3c_2}{2}f_x\eta_{xx}+c_2f\eta_{xxx}+\frac{c_2}{2}f_{xx}
\eta_x-\frac{3c_1}{4}f\eta_x
-\frac{3c_1}{8}f_x\eta+\varsigma_1\eta_{xx}-\varsigma_2\mu_{xx}+\frac{1}{2}(a\gamma)_x,\!\!
\label{eq4.00}
\\
\frac{\delta H_2}{\delta a}=2\varsigma_3a+vf+\frac{1}{2}\gamma\eta_x-\varsigma_2f_{x},\nonumber
\\
\frac{\delta H_2}{\delta\mu}=\gamma_xf+\frac{1}{2}\gamma f_x-\frac{1}{2}v\eta_x+\varsigma_2\eta_{xx}
-2\varsigma_3\mu_x.\nonumber
\end{gather}
Under the special freezing point
\begin{gather*}
\hat{U}_0=\left(\frac{c_1}{2}dx^2,0,0,0,\left(c_2,0,\frac{c_5}{2}\right)\right)\in\mathcal{G}_{\rm reg}^{*},
\end{gather*}
the system~\eqref{eq2.10} reads
\begin{gather}
u_t=c_2\left(\frac{\delta H_2}{\delta u}\right)_{xxx}-c_1\left(\frac{\delta H_2}{\delta u}\right)_{x},
\qquad
v_t=-c_5\left(\frac{\delta H_2}{\delta v}\right)_{x},\nonumber
\\
\psi_t=\frac{c_1}{4}\frac{\delta H_2}{\delta\psi}-c_2\left(\frac{\delta H_2}{\delta\psi}\right)_{xx},
\qquad
\gamma_t=c_5\frac{\delta H_2}{\delta\gamma}.
\label{eq4.2}
\end{gather}
Using~\eqref{eq4.0}, we have
\begin{gather*}
\frac{\delta H_2}{\delta u}=\Lambda^{-1}\left(\frac{\delta H_2}{\delta f}\right),
\qquad
\frac{\delta H_2}{\delta\psi}=\Pi^{-1}\left(\frac{\delta H_2}{\delta\eta}\right),
\qquad
c_5\frac{\delta H_2}{\delta v}=\frac{\delta H_2}{\delta a},
\qquad
c_5\frac{\delta H_2}{\delta\gamma}=-\frac{\delta H_2}{\delta\mu}.
\end{gather*}
The system~\eqref{eq4.2} becomes
\begin{gather*}
u_t=-\left(\frac{\delta H_2}{\delta f}\right)_{x},
\qquad
\psi_t=-\frac{\delta H_2}{\delta\eta},
\qquad
v_t=-\left(\frac{\delta H_2}{\delta a}\right)_{x},
\qquad
\gamma_t=-\frac{\delta H_2}{\delta\mu},
\end{gather*}
which is the desired system~\eqref{eq4.1} due to~\eqref{eq4.0} and~\eqref{eq4.00}.
We thus complete the proof of the theorem.
\end{proof}

\subsection{Examples}
\begin{ex}[an $L^2$-type metric $M_{1,0,\frac{1}{4},0,1,-1}$] The systems~\eqref{eq4.1} reduces to
\begin{gather}
f_t=\varsigma_1f_{xxx}-\varsigma_2a_{xx}-3ff_x+\frac{3}{8}\eta\eta_{xx}-aa_x+\frac{1}{2}\mu\mu_{xx},\nonumber
\\
\eta_t=4{\varsigma}_1\eta_{xxx}-3f\eta_x-\frac{3}{2}f_x\eta-4{\varsigma}_2\mu_{xx}-2(a\mu)_x,\nonumber
\\
a_t=\varsigma_2f_{xx}-2\varsigma_3a_x-(af)_x+\frac{1}{2}(\mu\eta_x)_x,
\label{eq4.5}
\\
\mu_t=\varsigma_2\eta_{xx}-2\varsigma_3\mu_x-\frac{1}{2}a\eta_x-\mu_xf-\frac{1}{2}\mu f_x.
\nonumber
\end{gather}
We call this system~\eqref{eq4.5} to be a~Kuper-2KdV equation.
Especially, (1)~if we set $\eta=\mu=0$, we have
\begin{gather*}
f_t=\varsigma_1f_{xxx}-\varsigma_2a_{xx}-3ff_x-aa_x,
\qquad
a_t=\varsigma_2f_{xx}-2\varsigma_3a_x-(af)_x,
\end{gather*}
which is a~two-component generalization of the KdV equation with three parameters including the Ito
equation in~\cite{Ito1982} for $\varsigma_1\ne0$, $\varsigma_2=\varsigma_3=0$; (2)~if we set
$\varsigma_1=\frac{1}{2}$, $\varsigma_2=0$, $a=0$ and $\mu=0$, we have
\begin{gather*}
f_t=\frac{1}{2}f_{xxx}-3ff_x+\frac{3}{8}\eta\eta_{xx},
\qquad
\eta_t=2\eta_{xxx}-3f\eta_x-\frac{3}{2}f_x\eta,
\end{gather*}
which is the Kuper--KdV equation in~\cite{Kuper1984}.
\end{ex}

Let us remark that when we choose $\varsigma_1=\frac{1}{4}$ and $\varsigma_2=\varsigma_3=0$, up to
a~rescaling, the Kuper--2KdV equation~\eqref{eq4.5} is the super-Ito equation (equation~(4.14b)
in~\cite{Fordy1989}) proposed by M.~Antonowicz and A.P.~Fordy, which has three Hamiltonian structures.
According to our terminologies, we would like to call it the Kuper--Ito equation.
\begin{ex}[an $H^1$-type metric $M_{1,1,\frac{1}{4},1,1,-1}$] The systems~\eqref{eq4.1} reduces to
\begin{gather}
f_t-f_{xxt}=\varsigma_1f_{xxx}-\varsigma_2a_{xx}-3ff_x+2f_xf_{xx}+ff_{xxx}
+ \frac{3}{8}\eta\eta_{xx}+\frac{1}{2}\eta_x\eta_{xxx}-aa_x+\frac{1}{2}\mu\mu_{xx},\nonumber
\\
\eta_{xxt}-\frac{1}{4}\eta_t=\frac{3}{4}f\eta_x+\frac{3}{8}f_x\eta-f\eta_{xxx}-\frac{1}{2}f_{xx}
\eta_x-\frac{3}{2}f_x\eta_{xx}-{\varsigma}_1\eta_{xxx}
+ {\varsigma}_2\mu_{xx}-\frac{1}{2}(a\mu)_x,
\label{eq4.6}
\\
a_t=\varsigma_2f_{xx}-2\varsigma_3a_x-(af)_x+\frac{1}{2}(\mu\eta_x)_x,\nonumber
\\
\mu_t=\varsigma_2\eta_{xx}-2\varsigma_3\mu_x-\frac{1}{2}a\eta_x-\mu_xf-\frac{1}{2}\mu f_x.\nonumber
\end{gather}
We call this system~\eqref{eq4.6} to be a~Kuper--2CH equation.
Especially, (1) if we set $\varsigma_1=\varsigma_2=\varsigma_3=0$ and $\eta=\mu=0$, we have
\begin{gather*}
f_t-f_{xxt}=2f_xf_{xx}+ff_{xxx}-3ff_x-aa_x,
\qquad
a_t=-(af)_x,
\end{gather*}
which is the 2-CH equation in~\cite{CLZ2006, F2006}; (2) if by setting
$\varsigma_1=\varsigma_2=\varsigma_3=0$, $a=0$ and $\mu=0$, the system~\eqref{eq4.6} becomes
\begin{gather*}
f_t-f_{xxt}=ff_{xxx}+2f_xf_{xx}-3ff_x+\frac{3}{8}\eta\eta_{xx}+\frac{1}{2}\eta_x\eta_{xxx},
\\
\eta_{xxt}-\frac{1}{4}\eta_t=\frac{3}{4}f\eta_x+\frac{3}{8}f_x\eta-f\eta_{xxx}-\frac{1}{2}f_{xx}
\eta_x-\frac{3}{2}f_x\eta_{xx},
\end{gather*}
which is the {Kuper--CH} equation in~\cite{CD2010, ZZ2011}.
\end{ex}

\section[Supersymmetric Euler equations on $\mathcal{G}_{\rm reg}^{*}$]{Supersymmetric Euler equations on $\boldsymbol{\mathcal{G}_{\rm reg}^{*}}$}
\label{section4}

In this section, we want to discuss a~class of supersymmetric Euler equations on $\mathcal{G}^{*}$ asso\-cia\-ted
to a~special metric $M_{c_1,c_2,c_1,c_2,c_5,-c_5}$.
Moreover, we present a~class of supersymmetric and bihamiltonian Euler equations.

\subsection[Supersymmetric Euler equations on $\mathcal{G}_{\rm reg}^{*}$ w.r.t.\
$M_{c_1,c_2,c_1,c_2,c_5,-c_5}$]{Supersymmetric Euler equations on $\boldsymbol{\mathcal{G}_{\rm reg}^{*}}$ w.r.t.\
$\boldsymbol{M_{c_1,c_2,c_1,c_2,c_5,-c_5}}$}
In this case, we have
\begin{gather*}
c_1=c_3,
\qquad
c_2=c_4,
\qquad
c_6=-c_5.
\end{gather*}
By setting $\phi=\eta_x$ and $\alpha=\mu_x$, we obtain
\begin{gather*}
u=c_1f-c_2f_{xx},\qquad \psi=c_2\eta_{xx}-c_1\eta,\qquad v=c_5a,\qquad \gamma=-c_5\mu.
\end{gather*}

Let us def\/ine a~superderivative $D$ by $D=\p_\theta+\theta\p_x$ and introduce two superf\/ields
\begin{gather*}
\Phi=\eta+\theta f,
\qquad
\Omega=\mu+\theta a,
\end{gather*}
where $\theta$ is an odd coordinate.
A direct computation gives
\begin{thm}\label{thm5.1}
The Euler equation~\eqref{eq2.6} on $\mathcal{G}_{\rm reg}^{*}$ w.r.t.\ $M_{c_1,c_2,c_1,c_2,c_5,-c_5}$ is
invariant under the supersymmetric transformation
\begin{gather*}
\delta f=\theta\eta_x,
\qquad
\delta\eta=\theta f,
\qquad
\delta a=\theta\mu_x,
\qquad
\delta\mu=\theta a
\end{gather*}
and could be rewritten as
\begin{gather}
c_1\Phi_t-c_2D^4\Phi_{t}=\varsigma_1D^6\Phi-\frac{3}{2}c_1\big(\Phi D^3\Phi+D\Phi D^2\Phi\big)\nonumber
\\
\hphantom{c_1\Phi_t-c_2D^4\Phi_{t}=}{}
+c_2\left(D\Phi D^6\Phi+\frac{1}{2}D^2\Phi D^5\Phi+\frac{3}{2}D^3\Phi D^4\Phi\right)\nonumber
\\
\hphantom{c_1\Phi_t-c_2D^4\Phi_{t}=}{}
-\varsigma_2D^4\Omega+\frac{1}{2}c_5\big(D\Omega D^2\Omega+\Omega D^3\Omega\big),
\label{super2.16}
\\
c_5\Omega_t=\varsigma_2D^4\Phi-2\varsigma_3D^2\Omega-\frac{1}{2}
c_5\big(D\Omega D^2\Phi+2D^2\Omega D\Phi+\Omega D^3\Phi\big).\nonumber
\end{gather}
\end{thm}

\subsection{Examples}
\begin{ex}[another $L^2$-type metric $M_{1,0,1,0,1,-1}$]
The system \eqref{super2.16} reduces to
\begin{gather}
\Phi_t=\varsigma_1D^6\Phi-\frac{3}{2}\big(\Phi D^3\Phi+D\Phi D^2\Phi\big)-\varsigma_2D^4\Omega+\frac{1}{2}
\big(D\Omega D^2\Omega+\Omega D^3\Omega\big),
\nonumber
\\
\Omega_t=\varsigma_2D^4\Phi-2\varsigma_3D^2\Omega-\frac{1}{2}
\big(D\Omega D^2\Phi+2D^2\Omega D\Phi+\Omega D^3\Phi\big).
\label{eq5.4}
\end{gather}
We call this system \eqref{eq5.4} to be a super-2KdV equation. Especially,
(1)~if we set $\eta=\mu=0$,
we recover the two-component KdV equation again;
(2)~but if we choose
$\varsigma_1=\frac{1}{2}$, $\varsigma_2=\varsigma_3=0$ and $\Omega=0$, the system \eqref{eq5.4} becomes
\begin{gather*}
\Phi_t=\frac{1}{2}D^6\Phi-\frac{3}{2}\big(\Phi D^3\Phi+D\Phi D^2\Phi\big),
\end{gather*}
equivalently in componentwise forms,
\begin{gather*}
f_t=\frac{1}{2}f_{xxx}-3ff_{x}+\frac{3}{2}\eta\eta_{xx},
\qquad
\eta_t=\frac{1}{2}\eta_{xxx}-\frac{3}{2}(f\eta)_x,
\end{gather*}
which is the super-KdV equation in~\cite{Math1988}.
\end{ex}
\begin{ex}[another $H^1$-type metric $M_{1,1,1,1,1,-1}$]
The system \eqref{super2.16} reduces to
\begin{gather}
\Phi_t-D^4\Phi_{t}=\varsigma_1D^6\Phi-\frac{3}{2}(\Phi D^3\Phi+D\Phi D^2\Phi)+\left(D\Phi D^6\Phi+\frac{1}{2}
D^2\Phi D^5\Phi+\frac{3}{2}D^3\Phi D^4\Phi\right)\nonumber
\\
\hphantom{\Phi_t-D^4\Phi_{t}=}{}
-\varsigma_2D^4\Omega+\frac{1}{2}
\big(D\Omega D^2\Omega+\Omega D^3\Omega\big),
\label{eq5.5}
\\
\Omega_t=\varsigma_2D^4\Phi-2\varsigma_3D^2\Omega-\frac{1}{2}
\big(D\Omega D^2\Phi+2D^2\Omega D\Phi+\Omega D^3\Phi\big).\nonumber
\end{gather}
We call this system \eqref{eq5.5} to be a super-2CH equation. Especially,
(1)~if we set $\varsigma_1=\varsigma_2=\varsigma_3=0$ and $\eta=\mu=0$,
we obtain the 2-CH equation in~\cite{CLZ2006, F2006} again;
(2)~but if by setting $\varsigma_1=\varsigma_2=\varsigma_3=0$ and $\Omega=0$,
the system \eqref{eq5.5} becomes
\begin{gather*}
\Phi_t-D^4\Phi_{t}=\left(D\Phi D^6\Phi+\frac{1}{2}D^2\Phi D^5\Phi+\frac{3}{2}D^3\Phi D^4\Phi\right)-\frac{3}
{2}\big(\Phi D^3\Phi+D\Phi D^2\Phi\big),
\end{gather*}
which is the super-CH equation in~\cite{DS2001}.
\end{ex}

\subsection[Supersymmetric and bihamiltonian Euler equations w.r.t.\ $M_{0,c_2,0,c_2,
c_5,-c_5}$ on $\mathcal{G}_{\rm reg}^{*}$]{Supersymmetric and bihamiltonian Euler equations\\ w.r.t.\ $\boldsymbol{M_{0,c_2,0,c_2,
c_5,-c_5}}$ on $\boldsymbol{\mathcal{G}_{\rm reg}^{*}}$}

Let us combine with Theorem~\ref{thm3.1} and Theorem~\ref{thm5.1}, we have
\begin{thm}
The Euler equation on $\mathcal{G}_{\rm reg}^*$ w.r.t.\ the metric
$M_{0,c_2,0,c_2, c_5,-c_5}$ is supersymmetric and bihamiltonian.
\end{thm}

\begin{ex}[an $\dot{H}^1$-type metric $M_{0,1,0,1,1,-1}$]
The systems \eqref{super2.16} reduces to
\begin{gather}
-D^4\Phi_{t}=\varsigma_1D^6\Phi+\left(D\Phi D^6\Phi+\frac{1}{2}D^2\Phi D^5\Phi+\frac{3}{2}
D^3\Phi D^4\Phi\right)\nonumber
\\
\hphantom{-D^4\Phi_{t}=}{}
-\varsigma_2D^4\Omega+\frac{1}{2}\big(D\Omega D^2\Omega+\Omega D^3\Omega\big),
\label{eq5.7}
\\
\Omega_t=\varsigma_2D^4\Phi-2\varsigma_3D^2\Omega-\frac{1}{2}
\big(D\Omega D^2\Phi+2D^2\Omega D\Phi+\Omega D^3\Phi\big).\nonumber
\end{gather}
We call this system \eqref{eq5.7} to be a super-$2$HS equation. Especially, $(i)$ if we set
$\varsigma_1=\varsigma_2=\varsigma_3=0$ and $\eta=\mu=0$, we have
\begin{gather*}
-f_{xxt}=2f_xf_{xx}+ff_{xxx}-aa_x,
\qquad
a_t=-(af)_x,
\end{gather*}
which is a 2-HS equation in \cite{Zuo-2010-1}; $(ii)$ if by setting $\varsigma_1=\varsigma_2=\varsigma_3=0$
 and $\Omega=0$,
the system \eqref{eq5.7} becomes
\begin{gather*}
-D^4\Phi_{t}=D\Phi D^6\Phi+\frac{1}{2}D^2\Phi D^5\Phi+\frac{3}{2}D^3\Phi D^4\Phi,
\end{gather*}
which is the super-HS equation in~\cite{POP2003,JO2008}.
\end{ex}

\section{Concluding remarks}

We have described Euler equations associated to
the GNS algebra and shown that under which conditions
there are superymmetric or bihamiltonian. Here we only obtain
some suf\/f\/icient conditions but not necessary conditions.
As an application, we have naturally presented several generalizations
of some well-known integrable systems including
the Ito equation, the 2-CH equation and the 2-HS equation. It is well-known that
the Virasoro algebra, the extended Virasoro algebra and the Neveu--Schwarz algebras
are subalgebras of the GNS algebra. Thus our result could be regarded as a
generalization of that related to those subalgebras, see for instances
\cite{Ara2007,AK1998,
Con2007,Con2003,Con2006,DS2001,EM1970,Guha2008,Guha2000,Guha2006,K2005,BLG2008,
KM2003,KW2009,Kolev2007,JO2008,JO2009,Man1976,Marsden,Mish1970,M1998,OK1987,Ra1981,Zuo-2010-1}
and references therein. In the past twenty years,
in this subject it has grown in many dif\/ferent directions, please see \cite{KW2009}
and references therein.
Finally let us point out that in this paper all super-Hamiltonian operators are even.
Recently, in \cite{POP2003, Liu2010-1,Liu2010-2}, the classical Harry--Dym
equation is supersymmetrized in two ways, either by even supersymmetric
Hamiltonian operators or by odd supersymmetric Hamiltonian operators. Notice
that the HS equation is one of a member of negative Harry--Dym hierarchy.
It would be interesting to investigate whether the above point of view has an
extension to the odd supersymmetric integrable system, for instance, the odd HS equation.

\subsection*{Acknowledgements}

The author thanks Qing-Ping Liu and the anonymous referees
for valuable suggestions
and Qing Chen, Bumsig Kim and Youjin Zhang for the continued supports.
This work is partially supported by ``PCSIRT'' and the Fundamental Research
Funds for the Central Universities (WK0010000024) and NSFC(11271345)
and SRF for ROCS,SEM.

\pdfbookmark[1]{References}{ref}

\LastPageEnding

\end{document}